\documentstyle[editedvolume,epsfig,psfig]{crckapb} 

\begin{opening}
\title{The impact of Spanish astrophysics in the 1990's }

\author{S.F. Sanchez}
\author{C.R. Benn }
\institute{Isaac Newton Group of Telescopes\\
 Aprt. 321, 38700-S/C de La Palma,\\
 Spain}

\end{opening}

\runningtitle{Impact of Spanish astrophysics}

\begin{document}


\begin{abstract}

Spanish astronomy has grown rapidly during the 1990s.  The number of
papers published per year has doubled, to about 4\% of the world
total.  The number of high-impact papers, however, remains constant at
about 1\% of the world total.  Papers from France, Germany and Italy
also have lower impact than their numbers suggest, while for the US,
UK, Japan, Netherlands and Australia, the fraction of high-impact
papers is approximately the same as the fraction of all papers.  We
suggest a few possible explanations for this difference.

\end{abstract}

\section{Introduction}

Professional astrophysics has a short tradition in Spain compared
with 
other European countries. However, the inauguration of two world-class 
observatories (La Palma, Calar Alto) 
in Spain in the 1980s has
allowed Spanish astronomy to develop rapidly in recent years,
with a large increase in the number of astronomers, in the number
of institutions, in public interest, and in funding.

We have made a quantitative
study of the productivity of Spanish astronomy
during the 1990s, in order to compare its standing with that
of astronomy in other countries.

\section{Growth of Spanish astronmy}
The number of papers published per year per astronomer is shown in Fig. 1a
for the two largest Spanish astronomical institutions
(IAC and IAA, which include 50\% of all Spanish astronomers),
and examples of two small institutions, one old 
(OAN, founded 1790, with 15 astronomers
in 1998), and one new 
(IFCA, founded 1994, with 12 astronomers in 1998).
The data are taken from lists of publications provided by the institutions
themselves.
Fig. 1a shows that there are no significant differences between
the productivities of astronomers at the different institutions,
and that the number of papers per astronomer has risen by a factor of two
between 1989 and 1998.
The number of Spanish astronomers as a function of time was deduced
from a variety of published statistics, including IAU membership figures
for 1993 and 1998.
The ratio between number of astronomers and gross domestic product 
(GDP) is now similar to that for most other countries active 
in astrophysics.
Combining the estimates of the number of papers per astronomer with 
the number of astronomers yields, as a function of time,
the estimated fraction of all 
astronomy papers worldwide published from Spain (Fig. 1b).
Our estimates of the Spanish fraction are similar to those
given on the ISI web pages.
This growth
over the last decade is larger than
for any of the other top 15 countries active in astrophysics.

\begin{figure}
\centerline{%
\begin{tabular}{c@{\hspace{0.01pc}}c}
\epsfysize=7cm
\epsffile{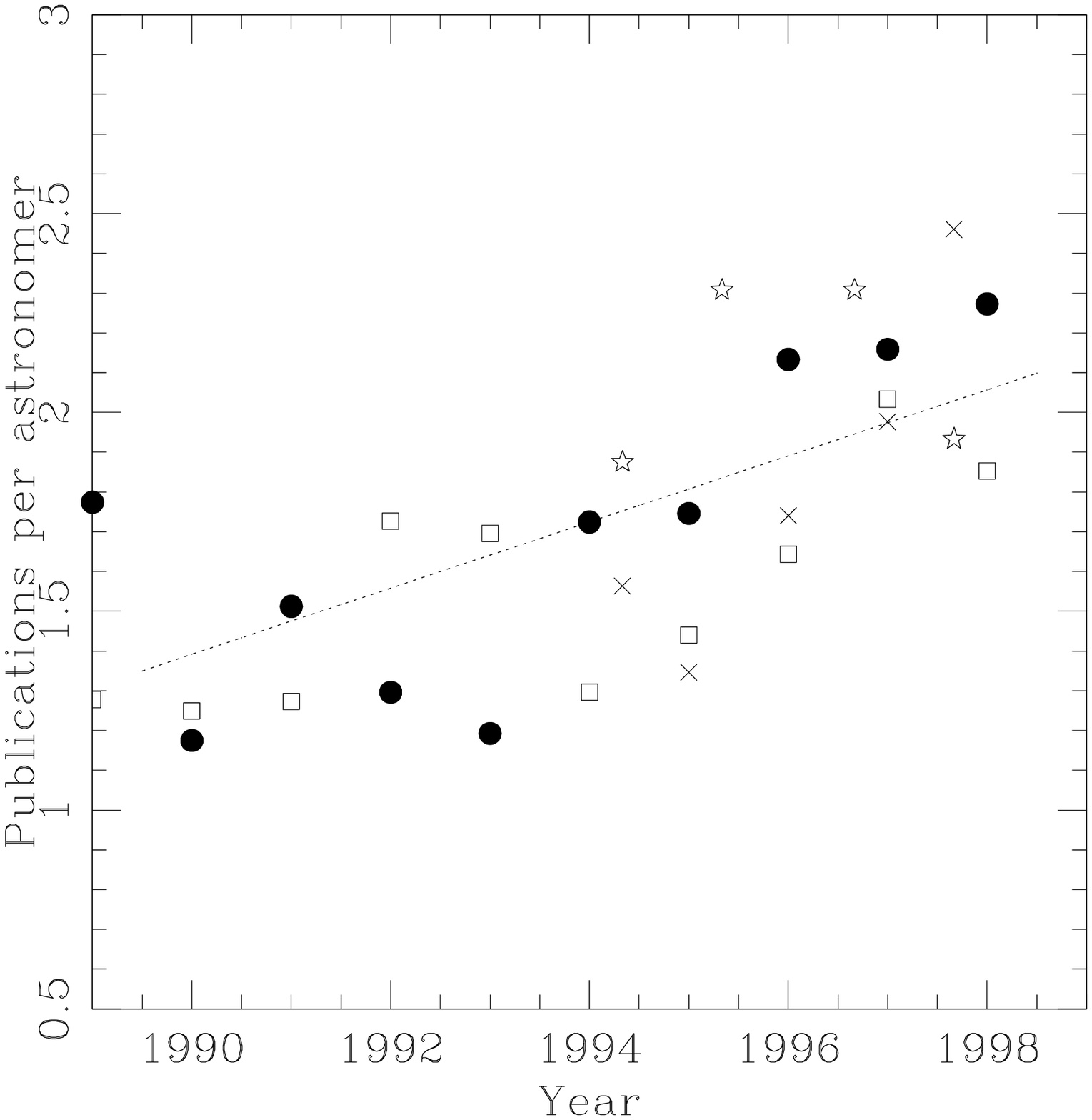} &
\epsfysize=7cm
\epsffile{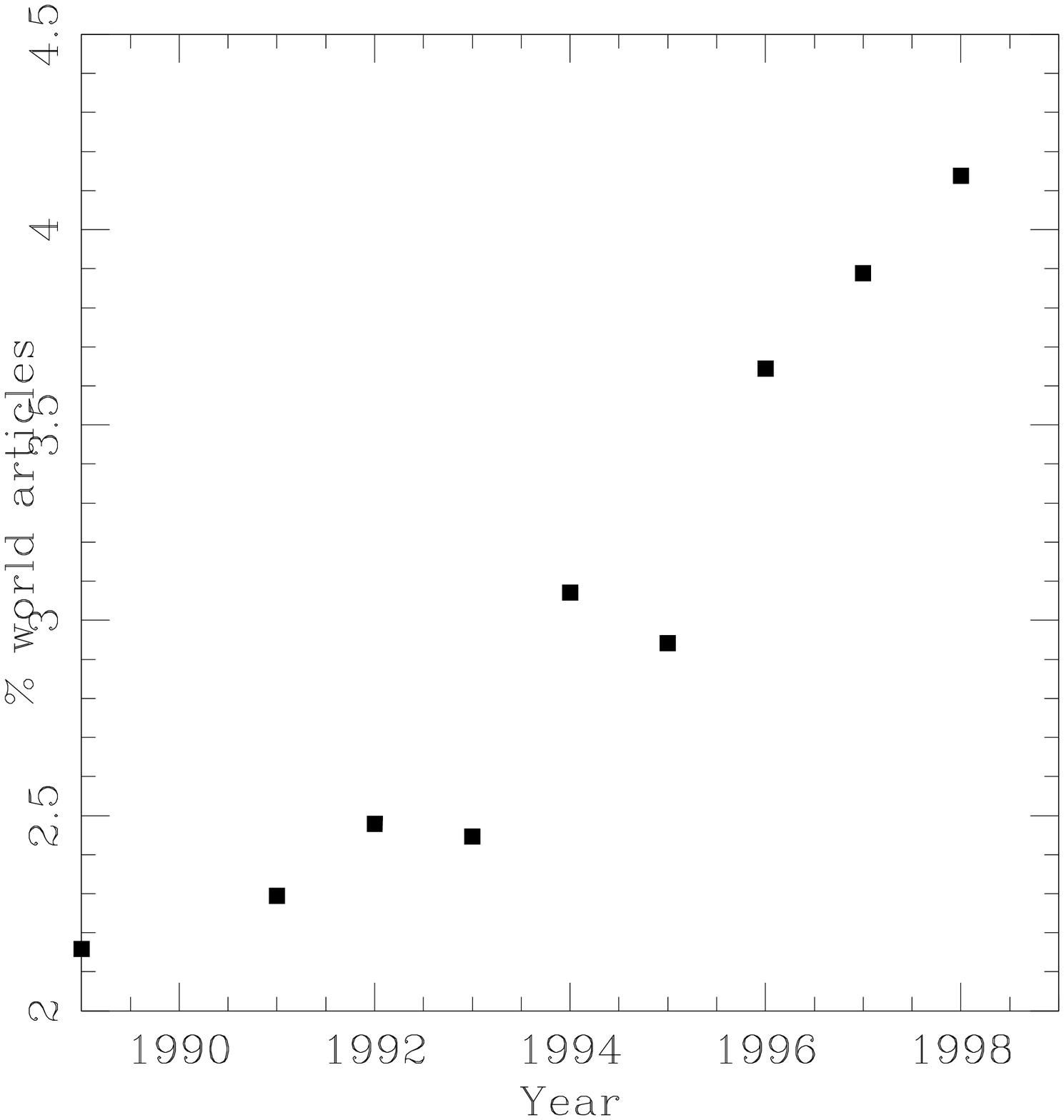} \\
\end{tabular}}
\caption{Growth of Spanish astronomy in the 1990s. (a) 
Number of papers published per astronomer in different
Spanish institutions: IAC (solid  circles), IAA
(squares), OAN
(crosses) and IFCA (stars). The solid line is the best-fit
straight line.
(b)
Fraction of astronomical papers published worldwide which originated
in Spain.
}

\label{fig:fig1}
\end{figure}

\begin{figure}
\centering
\psfig{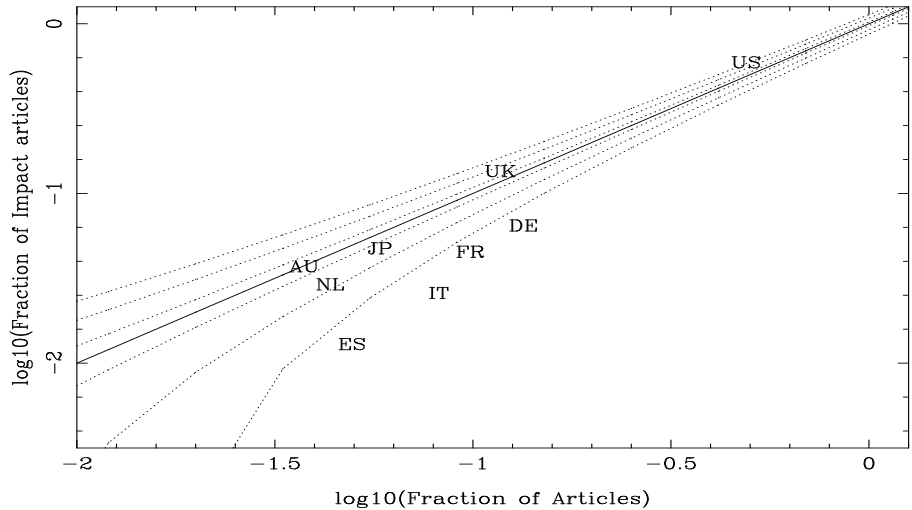}
\caption{
Scientific impact vs number of papers for different countries.
Both parameters are given as a fraction of the world total.
'Scientific impact' is an average of two measures, one based on the
number of citations to the top-1000 most-cited papers, one on the
number of papers published in Nature.
The solid line indicates \% impact = \% papers.
The dotted 
lines indicate 1, 3 and 5$\sigma$ departures from the solid line.
}
\label{fig:isi33}
\end{figure}

\section{Impact of Spanish astronomy}
As metrics of scientific productivity and impact, we counted for each country:

\begin{enumerate}

\item The number of papers published in refereed journals 
1994-99 (data taken from the web page of ISI, the
Institute of Scientific Information).

\item The number of citations to papers amongst the top-1000 most-cited
worldwide,
published during 1991-98.

\item The number of observational papers
published in Nature during 1989-98 (from a total of 453).
Nature has an impact factor (citations per paper) 8 times larger than
any astronomical journal.

\end{enumerate}
For each of the above counts, each paper was assigned to the country
hosting the first author.
Full details of the second and third datasets above, and 
a discussion of possible biases, are given by 
\citeauthor{bs00}\shortcite{bs00} and
\citeauthor{sb00}\shortcite{sb00}.

Number of papers published (the first metric)
is often used as a measure of scientific 
productivity, but may not reliably reflect scientific impact
(the last two metrics).
The average of the last two metrics is compared with the first,
in Fig. 2, 
for 9 countries for which data on all 
3 metrics were available.
For 5 of the countries (the US, UK, Japan, Netherlands and Australia),
the two fractions are similar.
For Germany, France, Italy and Spain, the ratio between impact and
total number of publications is a factor of $\sim$ 2.5 lower.
There are several possible explanations for the lower fraction of
high-impact papers in these counries.
One is language bias (which can affect citation 
rates in a number of ways);
in 4 of the 5 countries in the former
group, most scientific discourse and publication takes place in English.
Another possibility is that the level of resources provided affects
the fraction of high-quality papers.
In Spain and Italy, $\sim$ 1\% of the GDP is spent on research;
in most other countries (but including France and Germany), the 
fraction is
$\sim$ 2\%.  
A further possibility, highlighted by
\citeauthor{may97} \shortcite{may97}, 
is that the academic environment in the former group of countries
is more competitive, with most research done in universities rather
than dedicated institutions, and with more open competition for posts
(see Nature 1998 for discussion of the latter in Spain).

\section{Conclusions}
Spanish astronomy is a young science. The number of
professional astronomers, number of papers per astronomer and
fraction of papers (compared with the other countries)
have all risen by
a factor $\sim$2 in the last decade. 
The fraction of Spanish
astronomical papers has risen from $\sim$2\% in 1989 
a $\sim$4\% in 1998. However, the fraction of high-impact work has
remained constant at $\sim$1\% during this period.
For   
Spain (and for France, Germany and Italy) the ratio between the number of
high-impact papers, and the total number of papers published, is
lower than for the US, UK, Japan, Netherlands and Australia.
Possible reasons are language bias, smaller research budgets,
or less competitive deployment of research resources.

\end{document}